\documentclass[preprint,
showpacs,
 amsmath,amssymb,
 aps,
 prf,
]{revtex4-2}

\usepackage{amssymb}
\usepackage{amsthm}
\usepackage{graphicx}
\usepackage{dcolumn}
\usepackage{bm}
\usepackage{amsfonts}
\usepackage{amsmath}
\usepackage{amssymb}
\usepackage{color}
\usepackage{natbib}

\begin{document}
\title{Onsager reciprocal relations and chemo-mechanical coupling for chemically-active colloids}

\author{Marco De Corato}
\email{mdecorato@unizar.es}
\affiliation{Aragon Institute of Engineering Research (I3A), University of Zaragoza, Zaragoza, Spain\looseness=-1}

\author{Ignacio Pagonabarraga}
\affiliation{Departament de F\'{i}sica de la Mat\`{e}ria Condensada, Universitat de Barcelona, C. Mart\'{i} Franqu\`{e}s 1, 08028 Barcelona, Spain \\
University of Barcelona Institute of Complex Systems (UBICS), Universitat de Barcelona, 08028 Barcelona, Spain \\
CECAM, Centre Européen de Calcul Atomique et Moléculaire, École Polytechnique Fédérale de Lasuanne (EPFL), Batochime, Avenue Forel 2,1015 Lausanne, Switzerland\looseness=-1}

\begin{abstract}
Similar to cells, bacteria, and other microorganisms, synthetic chemically-active colloids can harness the energy from their environment through a surface chemical reaction and use its energy to self-propel in fluidic environments. In this paper, we study the chemo-mechanical coupling that leads to the self-propulsion of chemically active colloids. The coupling between chemical reactions and momentum transport is a consequence of the Onsager reciprocal relations. They state that the velocity and the surface reaction rate are related to the mechanical and chemical affinities through a symmetric matrix. A consequence of the Onsager reciprocal relations is that, if a chemical reaction drives the motion of the colloid, then an external force generates a reaction rate. Here, we investigate the Onsager reciprocal relations for a spherical active colloid that catalyzes a reversible surface chemical reaction between two species. We solve the relevant transport equations using a perturbation expansion and numerical simulations to demonstrate the validity of the reciprocal relations around the equilibrium. Our results are consistent with previous studies and highlight the key role of solute advection in preserving the symmetry of the Onsager matrix. Finally, we show that the Onsager reciprocal relations break down around a nonequilibrium steady state, which has implications for the thermal fluctuations of the active colloids used in experiments.

\end{abstract}

\maketitle

\section{Introduction}
Synthetic active colloids are microscopic particles that harness a catalytic chemical reaction to self-propel \cite{paxton2004catalytic,howse2007self}. These synthetic particles display biological-like features in that they are able to turn the chemical energy available in the environment into motion like bacteria or eukaryote cells. However, since their surface can be functionalized and their surface chemistry can be controlled during the manufacturing process, they represent potential candidates for novel cancer therapies \cite{hortelao2018targeting,tang2020enzyme,hortelao2020monitoring,tang2020enzyme}, cargo transport \cite{baraban2012transport} or environmental remediation \cite{parmar2018micro}. Such promising applications have sparked the development of many different synthetic active particles that propel through different mechanisms \cite{eloul2020reactive,de2020self,gallino2018physics}. A common feature of synthetic active colloids is that to move in fluidic environments they operate out of equilibrium to convert chemical energy into mechanical stresses, potentially leading to spontaneous symmetry-breaking instabilities \cite{Michelin_2013,de2013self,izri2014self,Narinder_2018,De_Corato_2020,De_Corato_2021}. Therefore, their behavior can be understood using the framework of nonequilibrium thermodynamics. 

In a recent series of papers Gaspard, Kapral and coauthors showed using thermodynamics considerations that, close to equilibrium, the velocity and the reaction rate of a chemically active particle are linearly related to an external force and to the chemical affinity \cite{gaspard2017communication,gaspard2018fluctuating,huang2018dynamics}. This chemo-mechanical coupling originates from the Onsager reciprocal relations and implies that, if a reaction rate drives self-propulsion in a certain direction, then a force applied in that direction drives a reaction rate. A consequence of the Onsager reciprocal relations is that it is possible to use external forces to drive chemical reactions. Similar examples of chemo-mechanical coupling are very common in biological settings, for instance, the adsorbtion of protein on cell membranes can change their preferential curvature \cite{tozzi2019out} and forces are known to impact reaction rates as in the case of mechanophores \cite{makarov2016perspective} or enzymatic reactions \cite{gumpp2009triggering}. In their work, Gaspard and Kapral \cite{gaspard2017communication} demonstrated that such chemo-mechanical coupling is relevant also for synthetic active colloids that propel through chemical reactions but they did not discuss the physical mechanism responsible for it. 

On the other hand, the self-propulsion of chemically-active colloids has been successfully explained using the framework of self-phoresis, which uses thin boundary layer asymptotics \cite{golestanian2007designing,moran2017phoretic}. According to this approach, the surface reaction generates a gradient of reactants and products that interact through a short-ranged potential with the surface of the active colloid \cite{anderson1982motion,anderson1989colloid}. This mechanism results in the development of a phoretic slip velocity within a few nanometers of the particle surface, which, in turn, drives the motion of the active colloid. While this framework successfully explains how a surface reaction results in self-propulsion it is not clear how an external force can generate a reaction rate. In these studies, the advective transport of the reactant and product species is usually neglected, and the transport of species is solved independently of the velocity field. As a consequence, the reaction rate is decoupled from the flow field and the symmetry of the Onsager relations appears to be broken.

In this paper, we address this point by investigating the physical mechanism leading to the chemo-mechanical coupling highlighted by Gaspard and Kapraal \cite{gaspard2017communication}. To do so, we use integral relations, a perturbation expansion and numerical simulations. We show that by, solving the transport equations around a chemically active colloid, without assuming a short-ranged interaction potential \cite{sharifi2013diffusiophoretic}, we recover a symmetric Onsager matrix. Our analysis reveals that the advection of the reactant and product species, which is often neglected, is the physical mechanism leading to the  symmetry of the chemo-mechanical coupling discovered by Gaspard and Kapraal \cite{gaspard2017communication}. Consistently taking into account advection is crucial to preserve the symmetry of the Onsager reciprocal relations in the case of self-propelled chemically-active particles.

Finally, since many experiments are carried out far from thermodynamic equilibrium, we investigate the validity of the Onsager reciprocal relations around a nonequilibrium steady state. In this case, there is a net entropy production at steady state that breaks the detailed balance and the microreversibility of the molecular trajectories. This does not necessarily break the reciprocal relations because the fulfilment of the detailed balance implies the Onsager reciprocal relations but not vice-versa. In fact, there are some situations in which the Onsager reciprocal relations and fluctuation-dissipation relations hold around nonequilibrium steady states despite the breakdown of the detailed balance \cite{gabrielli1996onsager,gabrielli1999onsager,dal2019linear}. 

The paper is divided as follows. In section \ref{sec1}, we briefly recall the Onsager's reciprocal relations demonstrated by Gaspard and Kapraal in the case of a chemically-active colloid. In Sections \ref{sec2}-\ref{sec4}, we define the problem, and the governing equations and derive their dimensionless form. In Section \ref{sec5} we report the governing equations linearized around a generic steady-state. In Section \ref{sec6} we address the Onsager's reciprocal relations around equilibrium using perturbative analysis and numerical simulations. In Section \ref{sec7} we address the Onsager's reciprocal relations around a nonequilibrium steady state. Finally, Section \ref{sec8} contains conclusions and discussions.

\section{Onsager reciprocal relations for a chemically-active colloid}\label{sec1}
In a series of papers Gaspard, Kapral and coauthors \cite{gaspard2017communication,gaspard2018fluctuating,huang2018dynamics} showed that for small thermodynamic forces, i.e. in the linear response regime, the velocity of the active particle, $\boldsymbol{V}$, and the net reaction rate, $W$, are linearly related to the thermodynamic forces:
\begin{equation}\label{fullonsager}
\left(\begin{array}{c} \boldsymbol{V} \\ W \end{array}\right) = 
\left(\begin{array}{cc} D_{VF} & D_{VA} \, \boldsymbol{u} \\ D_{WF} \, \boldsymbol{u} & D_{WA} \end{array}\right) \cdot \left(\begin{array}{c} \frac{\boldsymbol{F}}{k_BT} \\ A_{\text{rxn}} \end{array}\right) \,\, ,
\end{equation}
where $D_{VF}$ is the translational diffusion coefficient, $D_{WA} $ is the reaction-diffusion coefficient and the coefficients that couple the velocity to the reaction rate, $D_{VA}$, and the reaction rate to the external force $D_{WF}$ are equal $D_{WF}=D_{VA}$. In Eq. \eqref{fullonsager}, the unit vector $\boldsymbol{u}$ determines the direction of motion induced by a nonzero chemical activity. Here, we consider an axisymmetric case, the unit vector, $\boldsymbol{u}$, coincides with the z-axis unit vector $\boldsymbol{u}= \boldsymbol{e}_z$ and the velocity is determined by its z-component $V$. The thermodynamic forces are given by the chemical affinity, $A_{\text{rxn}}$ and by the mechanical affinity, $\boldsymbol{F}/k_BT$, which need to be small for Eq. \eqref{fullonsager} to be valid. Without any loss of generality, we consider the external force $\boldsymbol{F}$ to be acting along the z-axis $\boldsymbol{F} = F \boldsymbol{e}_z$. 

The matrix that appears in Eq. \eqref{fullonsager} is called the Onsager matrix and, near the equilibrium, it must be symmetric and positive definite. The properties of the Onsager matrix are a cornerstone result of thermodynamics and follow from the microscopic reversibility of the molecular trajectories at equilibrium. The application of the Onsager's reciprocal relations to the case of a self-propelled chemically active colloid implies that, if a nonzero chemical affinity leads to the motion of a colloid along the z-axis, then an external force directed along the z-axis results in a reaction rate \cite{gaspard2017communication}. While the Onsager's reciprocal relations are rigorously derived near equilibrium, there are instances where they hold also when the linearization is performed around a nonequilibrium steady state \cite{gabrielli1996onsager,gabrielli1999onsager,dal2019linear}. In what follows, we show that the Onsager's reciprocal relations are valid around equilibrium using a perturbative expansion and numerical simulations. Numerical simulations show that the reciprocal relations are broken around a nonequilibrium steady state.

\section{Problem definition}\label{sec2}
\begin{figure}[h!]
\centering
\includegraphics[width=0.7\textwidth]{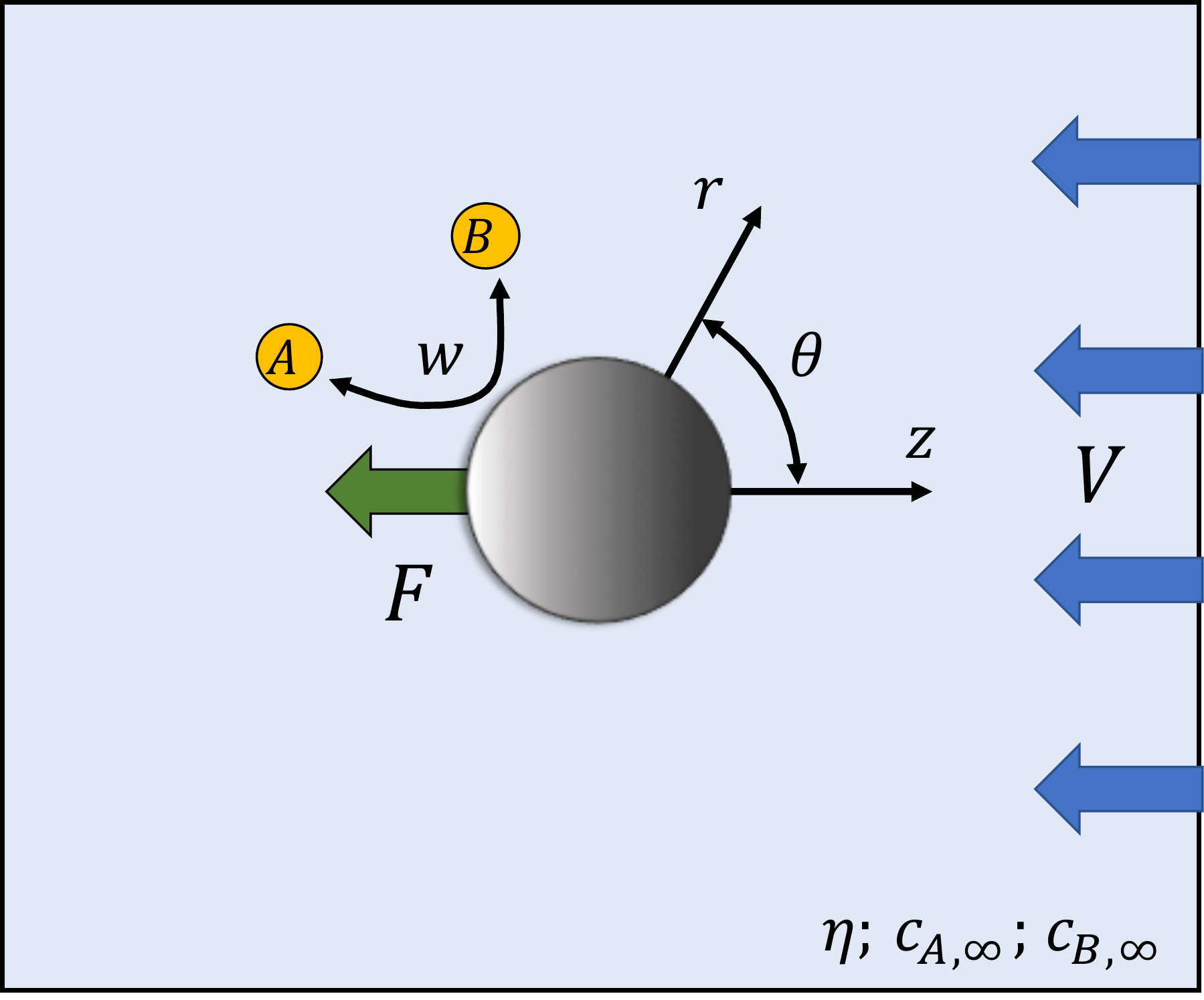}
\caption{Schematics of the system investigated. A chemically-active colloid is suspended in an incompressible fluid and a chemical reaction between two solute species, A and B, is catalyzed at the surface of the colloid. An external force might be acting along the z-axis. The concentration of the species A and B are fixed far from the colloid. An inhomogeneous interaction between the B solute molecules and the colloid surface drives the self-propulsion of the active particle. }
\label{fig_schem}
\end{figure}

To investigate the validity of the Onsager's reciprocal relations for an axisymmetric chemically active colloid, we study a thermodynamic system similar to that analyzed by Sabass and Seifert \cite{sabass2012dynamics} and depicted schematically in Figure \ref{fig_schem}. We consider an isothermal system comprising of a spherical particle of radius $R$ suspended in a dilute solution of two neutral species A and B whose chemical potentials are given by:
\begin{equation}\label{chempota}
\mu_A= k_B T \ln{c_A} \,\, ,
\end{equation}
\begin{equation}\label{chempotb}
\mu_B= k_B T \ln{c_B}+\Phi(r,\theta) \,\, ,
\end{equation}
with $c_A$ and $c_B$ the number density of species A and B and $k_B$ the Boltzmann constant and $T$ the absolute temperature. We assume that the chemical potential of the two species differs because of the interaction of the species B with the wall through the potential $\Phi(r,\theta) $ with $r$ and $\theta$ the radial and polar coordinates of a spherical coordinate system fixed at the particle center. 
We assume that the equilibrium reaction A$\rightleftharpoons$B takes place at the surface of the colloid according to the reaction rate per unit surface \cite{pagonabarraga1997fluctuating,bedeaux2011concentration}:
\begin{equation}\label{reacrate}
w = L_r (\theta) \left(1-\exp{\left(\frac{\mu_A-\mu_B}{k_B T}\right)}\right) \, \, \, \text{at} \, \, \, r=R \, \, ,
\end{equation}
with $L_r (\theta)$ the Onsager's coefficient that relates the local chemical potential to the local reaction rate. The total reaction rate, $W$ is given by the integral of $w$ over the active particle surface
\begin{equation}
W= \int_{S} w \,  \, dS \, \, ,
\end{equation}
with $S$ the surface of the particle.
To model chemically active colloids that are used in the experiments, which are coated with a catalyst on only some part of their surface, we consider that the reactivity changes along the particle surface as $L_r (\theta) = L_r \, g(\theta)$, where $g(\theta)$ is a positive dimensionless function and $L_r$ specifies the magnitude of the Onsager coefficient. 
To achieve self-propelled motion, the spherical symmetry of the problem needs to be broken \cite{golestanian2007designing,Uspal_2018,Burelbach_2019,Poehnl_2021}, which happens if the potential energy, $\Phi(r,\theta)$ or the Onsager's coefficient $L_r (\theta)$, changes along the polar angle. Here, we consider a potential energy that has the form $\Phi(r,\theta) = \Phi_0 f(r, \theta)$ with a characteristic magnitude $\Phi_0$ and varying in space according to the dimensionless function $f(r, \theta)$, which we assume to be axisymmetric around the z-axis. It follows that the molecules of B interact preferentially with one side of the surface than the other. Finally, we assume that the interaction potential decays to zero at large distances from the surface of the colloid $\Phi(r,\theta) \rightarrow 0$ as $r \rightarrow \infty$.

At thermodynamic equilibrium all the fluxes vanish, the suspending fluid is quiescent, the chemical potential is uniform and the distribution of the species A and B are given by the Boltzmann distribution:
\begin{equation} \label{equilibriumca}
c_A=c_{A, eq} = \text{const.} \, \, , 
\end{equation}
and 
\begin{equation} \label{equilibriumcb}
c_B=c_{B, eq}= c_{A, eq} \exp{\left(-\frac{\Phi_0}{k_BT} f(r,\theta)\right)} \, \, .
\end{equation}
For $r\rightarrow \infty$ the concentration of species A and B are equal because $\Phi(r,\theta)$ decays to zero. 

\section{General steady-state equations}\label{sec3}
By following the framework of nonequilibrium thermodynamics, we assume that the local thermodynamic forces and the local fluxes are linearly related even if the system is globally driven out of equilibrium \cite{de1962non}. 
We present the governing equations at steady state and in a reference frame attached to the center of the active particle. 
It follows that the momentum balance is given by,
\begin{equation}\label{mom_bal}
 \eta \boldsymbol{\nabla}^2 \boldsymbol{v} - \boldsymbol{\nabla} p = c_B \,\boldsymbol{\nabla} \mu_B +c_A \,\boldsymbol{\nabla} \mu_A \, \, ,
\end{equation}
where $\eta$ is the shear viscosity of the liquid, $\boldsymbol{v}$ is the velocity field and $p$ is the pressure. We neglected the inertia of the liquid in Eq. \eqref{mom_bal}, which is typically negligible at the colloidal scale.
By substituting the expression for the chemical potential $\mu_A$ and $\mu_B$, given by Eqs. \eqref{chempota}-\eqref{chempotb} in the momentum balance, we obtain
\begin{equation}
 \eta \boldsymbol{\nabla}^2 \boldsymbol{v} - \boldsymbol{\nabla} P = c_B \,\boldsymbol{\nabla} \Phi \, \, ,
\end{equation}
where we have defined the pressure $P$ as the sum of the hydrodynamic pressure and of the osmotic pressure $P= p+k_BT \, \left(c_A+c_B\right)$.
We assume that the fluid mixture is incompressible, therefore the continuity equation is given by
\begin{equation}
\boldsymbol{\nabla} \cdot \boldsymbol{v} = 0 \, \, ,
\end{equation}
with boundary conditions at infinity $r\rightarrow \infty$ given by:
\begin{equation}
\boldsymbol{v} = -V \boldsymbol{e}_z \, \, ,
\end{equation}
and at the surface of the particle $r=R$:
\begin{equation}
\boldsymbol{v} = \boldsymbol{0} \, \, .
\end{equation}
The balance of force on the active particle reads
\begin{equation}\label{force_bal}
\int_S \boldsymbol{T} \cdot \boldsymbol{n} \, dS = -\int_\Omega  c_B  \, \boldsymbol{\nabla} \Phi \, d\Omega - F \boldsymbol{e}_z\, \, ,
\end{equation}
where $\boldsymbol{T}$ is the stress tensor defined as $\boldsymbol{T}= \eta \left(\boldsymbol{\nabla}\boldsymbol{v}+\boldsymbol{\nabla}\boldsymbol{v}^T\right) - P \boldsymbol{I}$, $\boldsymbol{n}$ is the normal to the particle surface pointing into the fluid, and $\Omega$ is the volume outside the sphere.

The steady-state balance of the species A and B is given by
\begin{equation}
\boldsymbol{\nabla} \cdot \, \boldsymbol{J_A} =\boldsymbol{\nabla}  \cdot \, \boldsymbol{J_B} =0 \, \, ,
\end{equation}
where $\boldsymbol{J_A}$ and $\boldsymbol{J_B}$ are the fluxes of the species A and B, defined as
\begin{equation}\label{difffluxA}
\boldsymbol{J_A} = -\frac{L_{AA}}{T}\boldsymbol{\nabla} \mu_A +c_A \boldsymbol{v}\, \, ,
\end{equation}
\begin{equation}\label{difffluxB}
\boldsymbol{J_B} = -\frac{L_{BB}}{T}\boldsymbol{\nabla} \mu_B +c_B \boldsymbol{v}\, \, .
\end{equation}
The coefficients $L_{AA}$ and $L_{BB}$ are the Onsager's transport coefficients. The transport coefficients are related to the diffusion coefficients of the species A and B through $L_{AA}=D_A   \,T  \, c_A$ and $L_{BB}=D_B \, T \, c_B$, with $D_A$ and $D_B$ the diffusion coefficients of species A and B. In the definition of the diffusive fluxes, we have neglected the cross-coupling coefficients because we are considering dilute species. Nevertheless, the conclusions of the present work should hold in the case of cross diffusing species. 

At the surface of the active particle $r=R$, the fluxes of species A and B are related to the local reaction rate $w$, given by Eq. \eqref{reacrate}, and read
\begin{equation}
\boldsymbol{J_A}  \cdot \boldsymbol{n}  = -w\, \, ,
\end{equation}
\begin{equation}
\boldsymbol{J_B}  \cdot \boldsymbol{n}  = w\, \, .
\end{equation}
The net reaction rate is obtained by integrating $w$ over the surface of the active particle,
\begin{equation}
W = \int_S w \, dS \, \, .
\end{equation}
Far from the particle, $r\rightarrow \infty $, the chemical potential of species A is fixed, while the chemical potential of B is at equilibrium:
\begin{equation}
\mu_A \rightarrow \mu_{A, \infty}\, \, ,
\end{equation}
\begin{equation}
\mu_B    \rightarrow  \mu_{B,eq}\, \, .
\end{equation}
The difference between the chemical potential of the two species, normalized by $k_B T$, defines the chemical affinity, which is the driving force of the chemical reaction at the surface of the active particle. We define the chemical affinity, $A_{\text{rxn}}$, using the chemical potential of the species far from the particle
\begin{equation}\label{chemical_affinity}
A_{\text{rxn}}= \frac{\left(\mu_{A, \infty}-\mu_{B, eq} \right)}{k_B T} \, \, ,
\end{equation}
which is typically how the reaction rate is driven in experimental systems.
The thermodynamic forces that drive the active particle out of equilibrium are given by the mechanical affinity $F/k_BT$, acting directly on the particle, and by the chemical affinity, $A_{\text{rxn}}$ that drives the chemical reaction.

\section{Dimensionless equations}\label{sec4}
We make the governing equations dimensionless by using the following characteristic quantities: 
\begin{equation}
r= R \, \tilde{r} ; \, \, \, \, \boldsymbol{v} = \frac{k_B T \, R \, c_{A,eq}}{\eta } \, \tilde{\boldsymbol{v}}; \, \, \, \, P = k_B T \, c_{A,eq} \, \tilde{P}; \, \, \, \, c_{A}= c_{A,eq} \,  \tilde{c}_{A}; \, \, \, \, c_{B}= c_{A,eq} \, \tilde{c}_{B} \, \, .
\end{equation}
In the rest of the paper, we will consider dimensionless quantities only, and we omit the tilde superscript for clarity.
The dimensionless momentum balance reads:
\begin{equation}\label{dimensionlessmombal}
\boldsymbol{\nabla}^2 \boldsymbol{v} - \boldsymbol{\nabla} P = \epsilon \, c_B \,\boldsymbol{\nabla} f(r,\theta) \, \, ,
\end{equation}
with $\epsilon = \Phi_0/k_B T$ the dimensionless characteristic potential energy between the species B and the surface of the particle.
The mass balance reads:
\begin{equation}
\boldsymbol{\nabla} \cdot \boldsymbol{v} =  0 \, \, ,
\end{equation}
with boundary conditions at infinity $r\rightarrow \infty$ given by:
\begin{equation}
\boldsymbol{v} = -V \, \boldsymbol{e}_z \, \, ,
\end{equation}
and at the surface of the particle $r=1$:
\begin{equation}
\boldsymbol{v} = \boldsymbol{0} \, \, .
\end{equation}

The dimensionless balance of number density of species A and B is given by:
\begin{equation}
\boldsymbol{\nabla}^2 \, c_A  -\frac{Pe}{\beta} \, \boldsymbol{v} \cdot \boldsymbol{\nabla}c_A = 0 \, \, ,
\end{equation}
\begin{equation}
\boldsymbol{\nabla}^2 \,  c_B + \epsilon \, \boldsymbol{\nabla} \cdot \left[ c_B \, \boldsymbol{\nabla} f(r,\theta) \right] - Pe \, \boldsymbol{v} \cdot \boldsymbol{\nabla} c_B= 0 \, \, .
\end{equation}
Where the P\'eclet number is defined as $Pe = k_BT c_{A,eq} R^2 / \eta D_B $.  In defining the P\'eclet number we considered as characteristic velocity the one generated by the solute-surface interactions rather than the velocity of the particle.  This choice is dictated by the fact that the velocity of the active particle is unknown and is obtained from the solution of the equations.  
Alternatively, another velocity scale could be constructed using the external force $F$ but this choice results in the mechanical affinity being included in the P\'eclet number. As a consequence, one could not decouple the effects of an external force from the effects of advection.

In the limit $r \rightarrow \infty$, the chemical potential of the species A is kept to a constant value, which fixes their number density 
\begin{equation} 
c_{A} \rightarrow c_{A,\infty} \, \, .
\end{equation}
It is a nonzero chemical affinity that drives the reaction out of equilibrium. 
The number density of B decays to its equilibrium value far from the particle,  $r \rightarrow \infty$:
\begin{equation}
c_{B} \rightarrow 1 \, \, . 
\end{equation}
The chemical affinity that drives the chemical reaction is given by
\begin{equation}
A_\text{rxn}= \ln{c_{A,\infty}} \, \, .
\end{equation}

At the surface of the particle, $r=1$, the species react according to the reversible reaction:
\begin{equation}
-\boldsymbol{\nabla} c_{A} \cdot \boldsymbol{n} = -Da \, g(\theta) \, \left( 1-\frac{c_B}{ c_A} \, \exp{\left(\epsilon \, f(1,\theta)\right)}\right)\, \, , 
\end{equation}
\begin{equation} 
-\left[\boldsymbol{\nabla} c_{B} + \epsilon \, c_B \, \boldsymbol{\nabla} f(1,\theta)\right]\cdot \boldsymbol{n} = Da \, \beta \, g(\theta) \, \left( 1-\frac{c_B}{ c_A} \, \exp{\left(\epsilon \, f(1,\theta)\right)}\right)\, \, .
\end{equation}
Where $Da= L_r R /D_A c_{A,eq}$ is the Damkh{\"o}ler number defined with the diffusion coefficient of the species A, and $\beta = D_A/D_B$ is the  ratio of the diffusion coefficient of the two species. 
The average reaction rate can be evaluated by averaging the net consumption of A over the particle surface $S$:
\begin{equation}
W = Da \, \int_S \, g(\theta) \, \left( 1-\frac{c_B}{ c_A} \, \exp{\left(\epsilon \, f(1,\theta)\right)}\right) \, dS \, \, .
\end{equation}
The particle is dragged by an external force along the z-axis. The dimensionless force balance on the particle gives:
\begin{equation}\label{force_bal_dim}
\int_S \boldsymbol{T} \cdot \boldsymbol{n} \, dS = -\epsilon \, \int_\Omega \, c_B \, \boldsymbol{\nabla} f(r,\theta) \, d\Omega - \frac{\beta}{Pe} \, F^* \, \boldsymbol{e}_z\, \, ,
\end{equation}
with the dimensionless force $F^*= F/\eta D_A$. 
In the present form, Eqs. \eqref{dimensionlessmombal}-\eqref{force_bal_dim} are nonlinear and they must be linearized to connect the velocity of the particle $V$ and the reaction rate $W$ to the thermodynamic forces through a linear relation. In the following section, we linearize Eqs. \eqref{dimensionlessmombal}-\eqref{force_bal_dim} around a generic steady state.

\section{Linearization around a steady state}\label{sec5}
To derive the Onsager reciprocal relations derived directly from the transport equations, Eqs. \eqref{mom_bal}-\eqref{chemical_affinity}, we consider small deviations of the thermodynamic forces around their steady state value, $A_{\text{rxn}}=A_{\text{rxn},0}+\delta A_{\text{rxn}}$ and $F^* = F^*_0+ \delta F^*$, where $\delta A_{\text{rxn}}$ and $\delta F^*$ are small. We thus linearize the governing equations around the steady state. The number density of A and B, the velocity and the pressure fields are then expanded as $c_A=c_{A,0}+\delta c_A$, $c_B=c_{B,0}+\delta c_B$, $\boldsymbol{v}=\boldsymbol{v}_0+\delta \boldsymbol{v}$ and $P=P_0+\delta P$. Similarly, the velocity of the particle is given by $V=V_0+\delta V$ and the reaction rate by $W=W_0+\delta W$. The base state equations for the unknowns $c_{A,0}$, $c_{B,0}$, $\boldsymbol{v}_0$, $P_0$ $V_0$ and $W_0$ satisfy the same equations as Eqs. \eqref{dimensionlessmombal}-\eqref{force_bal_dim}. The equations for the deviation are obtained by substituting the expansions in the dimensionelss equations Eqs. \eqref{dimensionlessmombal}-\eqref{force_bal_dim} and neglecting the nonlinear terms. 
The linearized momentum and mass balance read:
\begin{equation}\label{lindimensionlessmombal}
 \eta \boldsymbol{\nabla}^2 \delta \boldsymbol{v} - \boldsymbol{\nabla} \delta P = \epsilon \, \delta c_B \,\boldsymbol{\nabla} f(r,\theta) \, \, ,
\end{equation}
\begin{equation}
\boldsymbol{\nabla} \cdot \delta \boldsymbol{v} = 0 \, \, .
\end{equation}
with boundary conditions at infinity $r \rightarrow \infty$ given by:
\begin{equation}
\delta \boldsymbol {v} = -\delta V \, \boldsymbol{e}_z \, \, ,
\end{equation}
and at the surface of the particle $r=1$:
\begin{equation}
\delta \boldsymbol{v} = \boldsymbol{0} \, \, .
\end{equation}
The force balance reads
\begin{equation}\label{force_bal1}
\int_S \left[ \left(\boldsymbol{\nabla}\delta\boldsymbol{v}+\boldsymbol{\nabla} \delta \boldsymbol{v}^T\right) - \delta P \, \boldsymbol{I} \right] \cdot \boldsymbol{n} \, dS = -\epsilon \, \int_\Omega \, \delta c_B  \, \boldsymbol{\nabla} f(r,\theta) \, d\Omega -  \frac{\beta}{Pe} \, \delta F^* \, \boldsymbol{e}_z\, \, .
\end{equation}

The linearized transport of the species A and B reads:
\begin{equation}
\boldsymbol{\nabla}^2  \delta c_A =0 \, \, ,
\end{equation}
\begin{equation}
\boldsymbol{\nabla}^2 \delta c_B + \epsilon \, \boldsymbol{\nabla} \cdot \left[ c_B \, \boldsymbol{\nabla} f(r,\theta) \right]- Pe \, \delta \boldsymbol{v} \cdot \boldsymbol{\nabla} c_{B,0} - Pe \,\boldsymbol{v}_0 \cdot \boldsymbol{\nabla} \delta c_{B}=0 \, \, .
\end{equation}
The reaction rate is also linearized, leading to the linearized boundary condition at $r=1$.
\begin{equation}
-\boldsymbol{\nabla} \delta c_{A} \cdot \boldsymbol{n} = Da \, g(\theta) \, \left( \frac{c_{A,0} \delta c_B-c_{B,0} \delta c_A}{ c_{A,0}^2} \right) \, \exp{\left(\epsilon \, f(1,\theta)\right)} \, \, , 
\end{equation}
\begin{equation} 
-\left[\boldsymbol{\nabla} \delta c_{B} + \epsilon \, \delta c_B \, \boldsymbol{\nabla} f(1,\theta)\right]\cdot \boldsymbol{n} = - Da \, \beta \, g(\theta) \, \left( \frac{c_{A,0} \delta c_B-c_{B,0} \delta c_A}{ c_{A,0}^2} \right) \, \exp{\left(\epsilon \, f(1,\theta)\right)}\, \, .
\end{equation}
The deviation of the concentration from the steady state far from the particle yields the boundary conditions:
\begin{equation}
\delta c_A \rightarrow  \delta c_{A, \infty} \, \, \text{as} \, \, r \rightarrow \infty,
\end{equation}
\begin{equation}\label{lindimensionlesscbboundcond}
\delta c_B \rightarrow 0 \, \, \text{as} \, \, r \rightarrow \infty .
\end{equation}
The deviation of the chemical affinity, $\delta A_{\text{rxn}}$, is related to the deviation of the far-field concentration, $\delta c_{A, \infty}$, through  $ \delta A_{\text{rxn}}= \delta c_{A, \infty}/ \exp{(A_{\text{rxn},0})}$, where $A_{\text{rxn},0}$ is the chemical affinity of the steady state around which the linearization is performed.

The reaction rate $\delta W$ can be evaluated by averaging the net consumption of A over the particle surface $S$:
\begin{equation}
\delta W = Da \, \int_S \, g(\theta) \, \left( \frac{c_{A,0} \delta c_B-c_{B,0} \delta c_A}{ c_{A,0}^2} \right) \, \exp{\left(\epsilon \, f(1,\theta)\right)} \, dS \, \, .
\end{equation} 
The velocity of the particle can be computed using the Lorentz reciprocal theorem \cite{masoud2019reciprocal}:
\begin{equation}\label{rectheorem}
\delta V = \frac{\beta}{6 \pi \, Pe} \, \delta F^*-\frac{\epsilon}{6 \pi} \, \int_\Omega \delta c_B \boldsymbol{\nabla} f(r,\theta) \cdot \hat{\boldsymbol{v}}_\text{Stokes} \, d\Omega \, \, ,
\end{equation}
where $\hat{\boldsymbol{v}}_\text{Stokes}$ is the Stokes flow past a sphere given by 
\begin{equation}
\hat{\boldsymbol{v}}_\text{Stokes} = \left(\frac{3}{2r}-\frac{1}{2r^3}-1 \right) \cos{(\theta)} \boldsymbol{e}_r-\left(\frac{3}{4r}+\frac{1}{4r^3}-1 \right) \sin{(\theta)} \boldsymbol{e}_\theta \,\, ,
\end{equation}
and $\boldsymbol{e}_r$ and $\boldsymbol{e}_\theta$ are the unit vectors along the radial and polar direction.

By linearizing the governing equations the deviation of the particle velocity, $\delta V$, and of the reaction rate, $\delta W$, are linearly related to the deviations of the thermodynamic forces as:
\begin{equation}\label{dimensionlessfullonsager}
\left(\begin{array}{c} \delta V \\ \delta W \end{array}\right) = 
\left(\begin{array}{cc} D_{VF} & D_{VA} \\ D_{WF}  & D_{WA} \end{array}\right) \cdot \left(\begin{array}{c} \frac{\beta}{Pe} \delta F^* \\ \delta A_{\text{rxn}} \end{array}\right) \,\, .
\end{equation}
To investigate the validity of the Onsager reciprocal relations, we are interested in calculating the cross-coupling coefficients $D_{VA}$ and $D_{WF}$ for a given steady state. To compute $D_{WF}$ we apply first an external force $\delta F^*$, we solve the system of equations given by Eqs \eqref{lindimensionlessmombal}-\eqref{lindimensionlesscbboundcond}, and we evaluate the reaction rate $\delta W$. The coefficient relating the applied force to the reaction rate is the Onsager coefficient $D_{WF}$. Likewise, to compute $D_{VA}$, we apply a chemical affinity $\delta A_{\text{rxn}}$, we solve the system of equations given by Eqs \eqref{lindimensionlessmombal}-\eqref{lindimensionlesscbboundcond}, and we calculate the particle velocity $\delta V$.

\section{Reciprocal relations around equilibrium}\label{sec6}
In the case of a base state given by the thermodynamic equilibrium, the Onsager's matrix given by Eq. \eqref{dimensionlessfullonsager} must be symmetric positive semi-definite. This property follows from the microscopic reversibility of the trajectories under time reversal. In what follows, we answer the question: in the case of a base state given by the thermodynamic equilibrium, do the transport equations, Eqs. \eqref{lindimensionlessmombal}-\eqref{lindimensionlesscbboundcond}, result in a symmetric positive semi-definite Onsager matrix? We address this question in the following subsections using a perturbation expansion and numerical simulations.
At thermodynamic equilibrium the base state is given by $c_{A,0}=1$, $c_{B,0}=\exp{(-\epsilon \, f(r,\theta) )}$, $\boldsymbol{v}_0=\boldsymbol{0}$, $P=k_BT \left(c_{A,0}+c_{B,0} \right)$, $V_0=0$ and $W_0=0$.

\subsection{Perturbation expansion for weak interaction potentials and small Damkh{\"o}ler numbers}
Even if the system of equation, given by Eqs \eqref{lindimensionlessmombal}-\eqref{lindimensionlesscbboundcond}, is linear, its analytical solution is complicated by the fact that the chemical activity and the potential energy vary with the polar angle $\theta$.  To circumvent this difficulty, we perform a perturbation expansion of the linearized equations, which is valid for small $\epsilon$ and small $Da$. 
\begin{alignat}{1}
& \delta\boldsymbol{v} = \delta \boldsymbol{v}^{0,0} + \epsilon \, \delta \boldsymbol{v}^{1,0}+  Da \, \delta \boldsymbol{v}^{0,1}+ \epsilon^2 \, \delta \boldsymbol{v}^{2,0} + \epsilon  \, Da \, \delta \boldsymbol{v}^{1,1}+ Da^2 \, \delta \boldsymbol{v}^{0,2}+\mathcal{O}(\epsilon^3, Da^2 \epsilon, \epsilon^2 Da, Da^3) \, \, ,\\
& \delta P = \delta P^{0,0} + \epsilon  \, \delta P^{1,0} + Da \, \delta P^{0,1}+ \epsilon^2 \, \delta P^{2,0} + \epsilon  \, Da \, \delta P^{1,1}+ Da^2 \, \delta P^{0,2}+\mathcal{O}(\epsilon^3, Da^2 \epsilon, \epsilon^2 Da, Da^3) \, \, ,\\
& \delta c_A =  \epsilon \, \delta c^{1,0}_A + Da \,\delta c^{0,1}_A+ \epsilon^2 \, \delta c^{2,0}_A + \epsilon  \, Da \, \delta c^{1,1}_A+ Da^2 \, \delta c^{0,2}_A+\mathcal{O}(\epsilon^3, Da^2 \epsilon, \epsilon^2 Da, Da^3) \, \, ,\\
& \delta c_B = \epsilon \, \delta c^{1,0}_B + Da \,\delta c^{0,1}_B+ \epsilon^2 \, \delta c^{2,0}_B + \epsilon  \, Da \, \delta c^{1,1}_B+ Da^2 \, \delta c^{0,2}_B +\mathcal{O}(\epsilon^3, Da^2 \epsilon, \epsilon^2 Da, Da^3) \, \, ,\\
& \delta W = \delta W^{0,0} + \epsilon  \, \delta W^{1,0} + Da \, \delta W^{0,1}+ \epsilon^2 \, \delta W^{2,0} + \epsilon  \, Da \, \delta W^{1,1}+ Da^2 \, \delta W^{0,2} +\mathcal{O}(\epsilon^3, Da^2 \epsilon, \epsilon^2 Da, Da^3) \, \, ,\\
& \delta V = \delta V^{0,0} + \epsilon  \, \delta V^{1,0} + Da \, \delta V^{0,1}+ \epsilon^2 \, \delta V^{2,0} + \epsilon  \, Da \, \delta V^{1,1}+ Da^2 \, \delta V^{0,2} +\mathcal{O}(\epsilon^3, Da^2 \epsilon, \epsilon^2 Da, Da^3) \, \, . 
\end{alignat}

Some of these terms can be shown to be zero based on simple considerations. The terms $Da \, \delta \boldsymbol{v}^{0,1}$, $Da^2 \, \delta \boldsymbol{v}^{0,2}$, $Da \, V_{0,1}$ and $Da^2 \, V^{0,2}$ are zero, because in the absence of a potential energy, $\epsilon=0$, the momentum balance is decoupled from the transport of mass and a reaction cannot generate fluid motion. Similarly, since the reaction rate is proportional to $Da$, there is no reaction rate if $Da=0$ and the terms $\delta W^{0,0}$, $\epsilon  \, \delta W^{1,0}$ and $\epsilon^2 \delta \, W^{2,0}$ are zero. 
In addition, we identify the field $\delta \boldsymbol{v}^{0,0}$ as the dimensionless Stokes flow past a sphere $\delta \boldsymbol{v}^{0,0}=\frac{\beta}{ Pe} F \hat{\boldsymbol{v}}_\text{Stokes}$ and the velocity $\delta V^{0,0}=\frac{\beta}{6 \pi \, Pe} F$.

With these simplifications in mind, the velocity of the active particle and the net reaction rate can be obtained from an expansion of the Onsager's matrix:
\begin{equation}\label{expanded_onsager}
\left(\begin{array}{c} \delta V \\ \delta W \end{array}\right) = 
\left(\begin{array}{cc} \frac{1}{6 \pi}+\epsilon \left( \, D^{1,0}_{VF}+\epsilon \,D^{2,0}_{VF}+ \, Da\,  D^{1,1}_{VF} \right)  & \epsilon \, Da \, D^{1,1}_{VF} \\ \epsilon \, Da\,D^{1,1}_{WF} & Da \left( \, D^{0,1}_{WA}+\epsilon \,  D^{1,1}_{WA}+Da \, D{0,2}_{WA} \right) \end{array}\right) \cdot \left(\begin{array}{c} \frac{\beta}{Pe} \delta F^* \\ \delta A_{\text{rxn}} \end{array}\right)  \,\, . 
\end{equation}
To leading order, the eigenvalues of the Onsager matrix, given by Eq. \eqref{expanded_onsager}, are $1/6\pi$ and $Da \,  D^{0,1}_{WA}$. Therefore, to demonstrate that the matrix is positive semi-definite, we need to show that $D^{0,1}_{WA} \geq 0$. To show that it is also symmetric, we need to prove that $D^{1,1}_{VA}=D{1,1}_{WF}$.
To do so, we plug the expansion into the governing equations above and solve order by order. The objective is to find the coefficient $D^{1,1}_{VA}$ that relates $\delta V$ and the chemical affinity, $\delta A_{rxn}$, and to show that it is equal to the coefficient $D^{1,1}_{WF}$. To do so, we proceed by dividing the problem into two steps. We first consider the case of a zero external force $\delta F^*=0$ and a nonzero chemical affinity $\delta A_{rxn}$, we calculate $\delta V^{1,1}$. The entry of the Onsager matrix $D^{1,1}_{VA}$ is simply given by the coefficient that relates  $\delta A_{rxn}$ and $\delta V^{1,1}$. We then we impose a nonzero $\delta F^*$ while keeping the chemical affinity to zero $\delta A_{rxn}=0$, we calculate $\delta W^{1,1}$ and obtain $D^{1,1}_{WF}$ as the coefficient that relates $\delta F^*$ and $\delta W^{1,1}$. 

The first-order reaction rate $\delta W^{1,1}$ and the velocity $\delta V^{1,1}$ are obtained using integral relations \cite{masoud2019reciprocal} that do not require the solution of all the fields.
By substituting the expansion in the governing equations, given by Eqs \eqref{dimensionlessmombal}-\eqref{rectheorem}, we find that the net reaction rate $\delta W^{1,1}$ is given by
\begin{equation}
\delta W^{1,1} = \int_S \, g(\theta) \,  \delta c^{1,0}_B  \, dS \, \, ,
\end{equation}
and that the velocity $\delta V{1,1}$ of the particle is given by
\begin{equation}\label{rectheorem1}
\delta V^{1,1} = -\frac{1}{6 \pi} \, \int_\Omega \, \delta c^{0,1}_B \, \boldsymbol{\nabla} f(r,\theta) \cdot \hat{\boldsymbol{v}}_\text{Stokes} \, d\Omega \, \, ,
\end{equation}
It follows that, to compute $\delta W^{1,1}$ and $\delta V^{1,1}$, we need to calculate the first order fields $\delta c^{1,0}_B$ and $\delta c^{0,1}_B$ only.

\subsubsection{Fixing the chemical affinity and calculating the particle velocity and reaction rate}
In order to find an expression for $\delta c^{0,1}_B$, we substitute the expansion in powers of $\epsilon$ and $Da$ and we keep all the terms linear in $Da$:
\begin{equation}\label{fixingchemaff1}
\boldsymbol{\nabla}^2 \, \delta c^{0,1}_B= 0 \, \, .
\end{equation}
with boundary condition at $r=1$ given by
\begin{equation}\label{fixingchemaff2}
-\boldsymbol{\nabla} \, \delta c^{0,1}_B \cdot \boldsymbol{n}= \beta \, g(\theta) \, \delta A_{\text{rxn}} \, \, .
\end{equation}
end with $\delta c^{0,1}_B=0$ as $r \rightarrow \infty$. The reaction rate, $\delta W^{0,1}$, is simply given by the integral of the reaction rate, given by Eq. \eqref{fixingchemaff2}, over the surface. This allows us to identify the coefficient $D^{0,1}_{WA}=\beta \, \int_S \, g(\theta) \, dS$. Since $\beta$ is always positive and $g(\theta)$ is a positive function, it follows that $D^{0,1}_{WA}\geq 0$, which proves that the Onsager's matrix is positive semi-definite.
 
The solution of the Eqs. \eqref{fixingchemaff1}-\eqref{fixingchemaff2} is obtained by expanding the distribution of the kinetic constant, $g(\theta)$, in Legendre polynomials as $g(\theta) = \sum_{l=0}^{\infty} g_l P_l(\cos{(\theta)})$, with $P_l$ the Legendre polynomial of order $l$.
The solution then reads
\begin{equation}
\delta c^{0,1}_B = \beta \, \delta A_{\text{rxn}} \, \sum_{l=0}^{\infty} \, \frac{g_l}{l+1} r^{-l-1} \,  P_l(\cos{(\theta)}) \, \, .
\end{equation}
Substituting this expression in the velocity we obtain 
\begin{equation}\label{rectheorem2}
\delta V^{1,1} = -\frac{\beta \,  \delta A_{\text{rxn}}}{6 \pi} \, \sum_{l=0}^{\infty} \, \frac{g_l}{l+1} \, \int_\Omega r^{-l-1} \,  P_l(\cos{(\theta)}) \boldsymbol{\nabla} f(r,\theta) \cdot \hat{\boldsymbol{v}}_\text{Stokes} \, d\Omega \, \, .
\end{equation}
Eq. \eqref{rectheorem2} allows us to identify the coefficient $D^{1,1}_{VA}$ as the proportionality constant between $\delta A_{\text{rxn}}$ and $\delta V^{1,1}$:
\begin{equation}\label{xi11}
D^{1,1}_{VA} = -\frac{\beta }{6 \pi} \, \sum_{l=0}^{\infty} \frac{g_l}{l+1} \int_\Omega \, r^{-l-1} \,  P_l(\cos{(\theta)}) \, \boldsymbol{\nabla} f(r,\theta) \cdot \hat{\boldsymbol{v}}_\text{Stokes} \, d\Omega \, \, .
\end{equation}

\subsubsection{Fixing the external force and calculating the reaction rate}
In order to find an expression for $\delta c^{1,0}_B$, we substitute the expansion in powers of $\epsilon$ and $Da$ and we keep all the terms linear in $\epsilon$:
\begin{equation}
\boldsymbol{\nabla}^2 \, \delta c^{1,0}_B= - \frac{\beta}{6 \pi } \, \delta F^* \, \boldsymbol{\nabla} f(r,\theta) \cdot \hat{\boldsymbol{v}}_\text{Stokes} \, \, .
\end{equation}
with boundary condition at $r=1$ given by
\begin{equation}
-\boldsymbol{\nabla} \, \delta c^{1,0}_B \cdot \boldsymbol{n}= 0 \, \, .
\end{equation}
and at infinity $\delta c^{1,0}_B=0$.
The second Green's theorem states that the following integral relation holds between $\delta c^{1,0}_B$ and an auxiliary field $\Psi$, which satisfies $\boldsymbol{\nabla}^2 \,\Psi =0$
\begin{equation}\label{greentheorem}
\frac{\beta}{6 \pi } \, \delta F^* \, \int_\Omega \Psi \boldsymbol{\nabla} f(r,\theta) \cdot \hat{\boldsymbol{v}}_\text{Stokes} \, d\Omega = \int_S \, \delta c^{1,0}_B \boldsymbol{\nabla} \Psi \cdot \boldsymbol{n} \, dS \, \, .
\end{equation}
Since the function $\Psi$ satisfies the Laplace equation, its solution can be written as $\Psi = \sum_{l=0}^{\infty} r^{-l-1} P_l(\cos{(\theta)})$, which we substitute in the expression above to obtain
\begin{equation}\label{greentheorem1}
-\frac{\beta}{6 \pi } \, \delta F^* \, \int_\Omega \sum_{l=0}^{\infty} \frac{r^{-l-1}}{l+1} P_l(\cos{(\theta)}) \boldsymbol{\nabla} f(r,\theta) \cdot \hat{\boldsymbol{v}}_\text{Stokes} \, d\Omega = \sum_{l=0}^{\infty} \int_S \delta c^{1,0}_B P_l(\cos{(\theta)}) \, dS \, \, .
\end{equation}
We now expand the function $\delta c^{1,0}_B$, evaluated at the surface of the colloid, in series of Legendre polynomials $\delta c^{1,0}_B = \sum_{l=0}^{\infty} \delta c^{1,0, l}_B \, P_l (\cos{(\theta)})$. We plug this expansion in the right hand side of Eq. \eqref{greentheorem1}, we apply the orthogonality property of the Legendre polynomials and equate term by term to get
\begin{equation}\label{greentheorem2}
-\frac{\beta}{6 \pi } \, \delta F^* \, \int_\Omega  \frac{r^{-l-1}}{l+1} P_l(\cos{(\theta)}) \boldsymbol{\nabla} f(r,\theta) \cdot \hat{\boldsymbol{v}}_\text{Stokes} \, d\Omega = \frac{2}{2l+1}\, \delta c^{1,0, l}_B\, \, .
\end{equation}
The equation above yields all the Legendre modes of the distribution $\delta c^{1,0, l}_B$ at the surface of the colloid. We can use this expression to evaluate the net reaction rate:
\begin{equation}
\delta W^{1,1} = \int_S g(\theta) \,  \delta c^{1,0}_B  \, dS \, \, ,
\end{equation}
where we now expand both $g(\theta)$ and $ \delta c^{1,0}_B$ in series of Legendre polynomials. By further using the orthogonality property of the Legendre polynomials, we obtain:
\begin{equation}
\delta W^{1,1} = \sum_{l=0}^{\infty} \frac{2}{2l+1} g_l \,  \delta c^{1,0, l}_B \, \, .
\end{equation}
we now substitute $\delta c_{B,1,0}^l$ obtained from Eq. \eqref{greentheorem2} to obtain:
\begin{equation}\label{finaleqreac}
\delta W^{1,1} = -\frac{\beta}{6 \pi } \, \delta F^* \, \sum_{l=0}^{\infty} \frac{g_l }{l+1} \int_\Omega r^{-l-1} \, P_l(\cos{(\theta)}) \boldsymbol{\nabla} f(r,\theta) \cdot \hat{\boldsymbol{v}}_\text{Stokes} \, d\Omega \, \, .
\end{equation}
Eq. \eqref{finaleqreac} relates the reaction rate to the mechanical affinity. The coefficient of proportionality between the reaction rate and the mechanical affinity yields the Onsager coefficient $D_{WF, 1,1}$, which is identical to that obtained in Eq. \eqref{xi11}:
\begin{equation}\label{finaleqreac1}
D^{ 1,1}_{WF} = -\frac{\beta}{6 \pi } \, \sum_{l=0}^{\infty} \frac{g_l }{l+1} \, \int_\Omega r^{-l-1}P_l(\cos{(\theta)}) \boldsymbol{\nabla} f(r,\theta) \cdot \hat{\boldsymbol{v}}_\text{Stokes} \, d\Omega = D^{ 1,1}_{VA}\, \, .
\end{equation}
This result also proves that, to leading order, the Onsager matrix given by Eq. \eqref{expanded_onsager} is symmetric for any choice of the distribution of the chemical activity, $g(\theta) \geq 0$ and for any choice of the distribution of the interaction energy $f(r,\theta)$. Interestingly, to leading order, neither $D_{WF}$ nor $D_{VA}$ depend on the P\'eclet number which suggests a negligible impact of advection to the cross-coupling coefficients. As a consequence, one would be tempted to neglect this mechanism when modeling chemically active colloids. Yet, neglecting \textit{a} \textit{priori} the transport due to advection in the diffusive fluxes, given by Eqs \eqref{difffluxA}-\eqref{difffluxB}, implies $D^{1,1}_{WF}=0$, thus breaking the symmetry of the Onsager matrix.
 
\subsubsection{Comparison of the self-diffusiophoretic velocity with previous results}
We can compare the self-diffusiophoretic velocity of the active colloid predicted by Eq. \eqref{rectheorem2} to that obtained by Sabbass and Seifert \cite{sabass2012dynamics}, in the limit of a short-range interaction potential, zero P\`{e}clet number and equal diffusivity of the two species A and B. The authors calculated the velocity of an active particle using a matched asymptotic expansion, which is valid for an interaction potential that decays quickly for $r>1$. In the case of an interaction potential that is only a function of the radius $\Phi(r)= \epsilon \, f(r)$, they find that the velocity depends on the dipolar mode of the reaction rate. Rewriting their result in dimensionless form and in the limit of slow reaction rate $Da \ll 1$ and weak interaction potentials $\epsilon \ll 1$:
\begin{equation}\label{previousstudy}
V_{dph} = -\frac{Da \, \epsilon \,  g_1 \, A_\text{rxn} \, \beta}{3} \int_1^{\infty} (r-1) f(r) \, dr \, \, .
\end{equation}
Where $Da \, g_1 \, A_\text{rxn}$ represents the dipolar component of the reaction rate occurring at the surface of the active colloid, and $f(r)$ is a quickly decaying function.
To compare with Eq. \eqref{previousstudy}, we rewrite Eq. \eqref{rectheorem2} for the case of the interaction potential being a function of the radial distance only $f(r,\theta)=f(r)$:
\begin{equation}\label{rectheorem3}
\delta V^{1,1} = -\frac{\beta \,  \delta A_{\text{rxn}}}{6 \pi} \, \sum_{l=0}^{\infty} \frac{g_l}{l+1} \int_\Omega r^{-l-1} \,  P_l(\cos{(\theta)}) \frac{\partial}{\partial r} f(r) 
\left(\frac{3}{2r}-\frac{1}{2r^3}-1 \right) \cos{(\theta)} \, d\Omega \, \, .
\end{equation}
We rewrite the integral above in spherical coordinates and carry out the integration along the azimuthal direction, which is trivial because the integrand does not depend on the azimuthal angle:
\begin{equation}\label{rectheorem4}
\delta V^{1,1} = -\frac{\beta \,  \delta A_{\text{rxn}}}{3} \, \sum_{l=0}^{\infty} \frac{g_l}{l+1} \int_1^{\infty} \int_0^{\pi} r^{-l+1} \, \sin{\theta} \,  P_l(\cos{(\theta)}) \frac{\partial}{\partial r} f(r) 
\left(\frac{3}{2r}-\frac{1}{2r^3}-1 \right) \cos{(\theta)} \, dr \, d\theta \, \, .
\end{equation}
We remove the radial derivative on the potential energy using integration by parts and we use the fact that $f(r) \rightarrow 0 $ as $r\rightarrow \infty$ and that  $\left(\frac{3}{2r}-\frac{1}{2r^3}-1 \right)=0$ at $r=1$ to obtain
\begin{equation}
\delta V^{1,1} = \frac{\beta \,  \delta A_{\text{rxn}}}{3} \sum_{l=0}^{\infty} \frac{g_l}{l+1} \int_1^{\infty} \int_0^{\pi} \, f(r) P_l(\cos{(\theta)}) \sin{\theta} \cos{(\theta)} \frac{\partial}{\partial r} \left[r^{-l+1}
\left(\frac{3}{2r}-\frac{1}{2r^3}-1 \right) \right] \, dr \, d\theta \, \, .
\end{equation}
We carry out the integral along the polar angle first. Since $\cos{(\theta)}=P_1(\cos{(\theta)})$, we can apply the orthogonality property of the Legendre polynomials, $\int_0^{\pi} P_l(\cos{(\theta)})P_{l'}(\cos{(\theta)}) \sin{(\theta)} d\theta = \delta_{ll'} 2/(2l+1)$, which identifies the mode $l=1$ as the only contribution in the summation:
\begin{equation}
\delta V{1,1} = \frac{\beta \, g_1 \, \delta A_{\text{rxn}}}{9} \, \int_1^{\infty}  \, f(r)
\left(\frac{3}{2r^4}-\frac{3}{2r^2} \right)  \, dr \, \, .
\end{equation}
We are left with an integration of the product between two functions along the radial coordinate . Since $f(r)$ decays quickly to zero, we can Taylor expand the term in the bracket around $r=1$ and we retain only the zeroth-order term \cite{perturbationmethods}. By doing this, we obtain the leading order propulsion velocity
\begin{equation}\label{sameresult}
\delta V = \frac{Da \, \beta \, \epsilon \,  g_1 \, \delta A_{\text{rxn}}}{3} \, \int_1^{\infty}  \, f(r)
\left(r-1 \right)  \, dr \, \, ,
\end{equation}
which for $\beta=1$ is exactly the same result as in Eq. \eqref{previousstudy}. Our results, which are derived from a model where the advective transport of species is considered, coincide with those where advection is neglected \cite{sabass2012dynamics}.  This suggests that, in the limit of a rapidly decaying interaction potential or weak interaction energy energy,  the advective transport of solute does not contribute to the propulsion velocity.   

\subsubsection{Onsager relations using numerical simulations around equilibrium}
We extend the perturbative analysis presented in the previous sections to non-vanishing values of $Da$ and $\epsilon$ by solving Eqs.\eqref{lindimensionlessmombal}-\eqref{lindimensionlesscbboundcond} using the finite element method. We consider the case of an asymmetric chemical activity given by $g(\theta)= 1+\cos{(\theta)}$ and an interaction potential that decays exponentially over a dimensionless lengthscale $\lambda^{-1}$ and is fore-aft asymmetric $f(r,\theta)= \epsilon \exp{[\lambda (r-1)]}\left(\cos{(\theta})-1\right)$. We further assume equal species diffusivity $\beta=1$. The computational domain is axisymmetric and it is divided into triangular elements, with a more refined mesh near the particle surface and coarser elements further away. To avoid finite size effects, the computational domain is chosen 500 times the radius of the active particle. A quadratic interpolation is used for the velocity field and the solute concentration fields and a linear interpolation is used for the pressure field. To derive the Onsager cross coupling coefficients, we proceed by fixing $\delta F^*_0 =1$ and $\delta A_\text{rxn}=0$ and evaluating the reaction rate we compute the coefficient $D_{WF}$. We then fix $\delta F^*_0 =0$ and $\delta A_\text{rxn}=1$ and evaluate the particle velocity we obtain the coefficient $D_{VA}$. 

\begin{figure}[h!]
\centering
\includegraphics[width=1.0\textwidth]{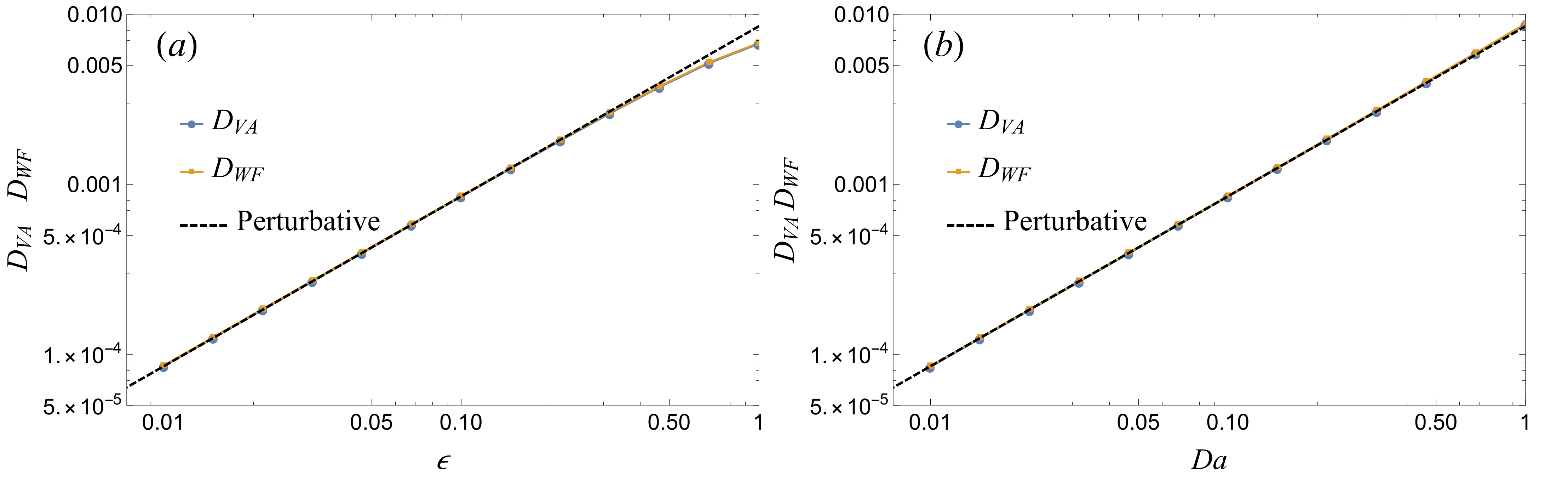}
\caption{Onsager cross coupling coefficients, $D_{WF}$ and $D_{VA}$, computed using numerical simulations of the governing equations linearized around the equilibrium. In the panel (a) we fix $Da=0.1$ and change $\epsilon$, while in panel (b) we fix $\epsilon =0.1$ and we change $Da$. The remaining dimensionless numbers are $\beta=1$, $\lambda=1$ and  $Pe=1$.}
\label{fig2}
\end{figure}

In Figure \ref{fig2}, we report the coefficients $D_{WF}$ and $D_{VA}$ for different values of $Da$ and $\epsilon$. In panel (a), the Onsager's coefficients, $D_{WF}$ and $D_{VA}$, are reported as a function of $\epsilon$ for $Da=0.1$ while, in panel (b), the coefficients are plotted against $Da$ for $\epsilon=0.1$. For the particular choice of parameters, the numerical results confirm the symmetry of the Onsager matrix and show that the perturbative approximation, given by Eq. \eqref{xi11} and Eq. \eqref{finaleqreac1}, are accurate for the cases shown in Figure \ref{fig2}.

\begin{figure}[h!]
\centering
\includegraphics[width=1.0\textwidth]{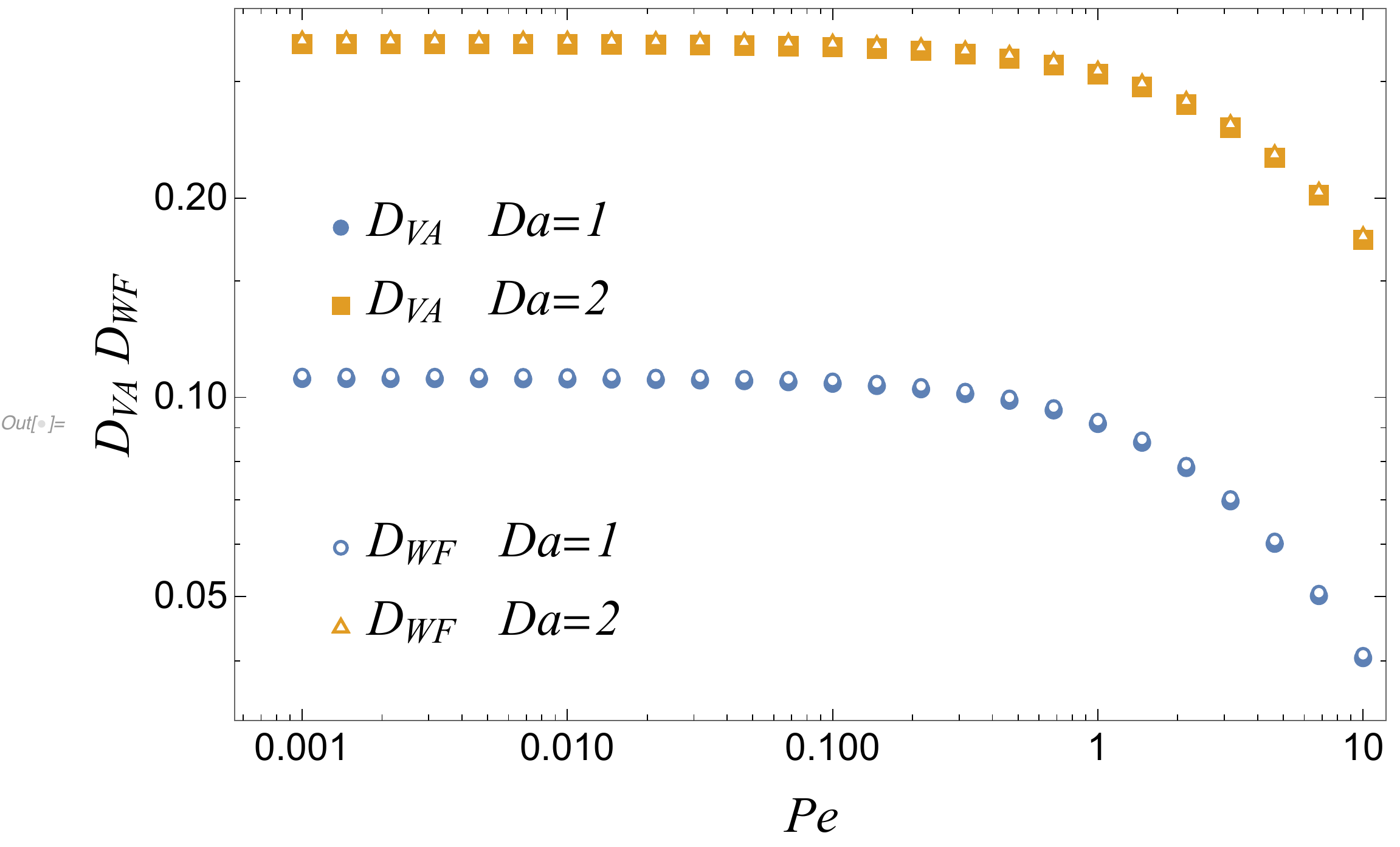}
\caption{Onsager cross coupling coefficients, $D_{WF}$ and $D_{VA}$, computed using numerical simulations of the governing equations linearized around the equilibrium.  The dimensionless numbers are $\beta=1$, $\lambda=1$ and  $\epsilon=1$.}
\label{fig4}
\end{figure}

\begin{figure}[h!]
\centering
\includegraphics[width=1.0\textwidth]{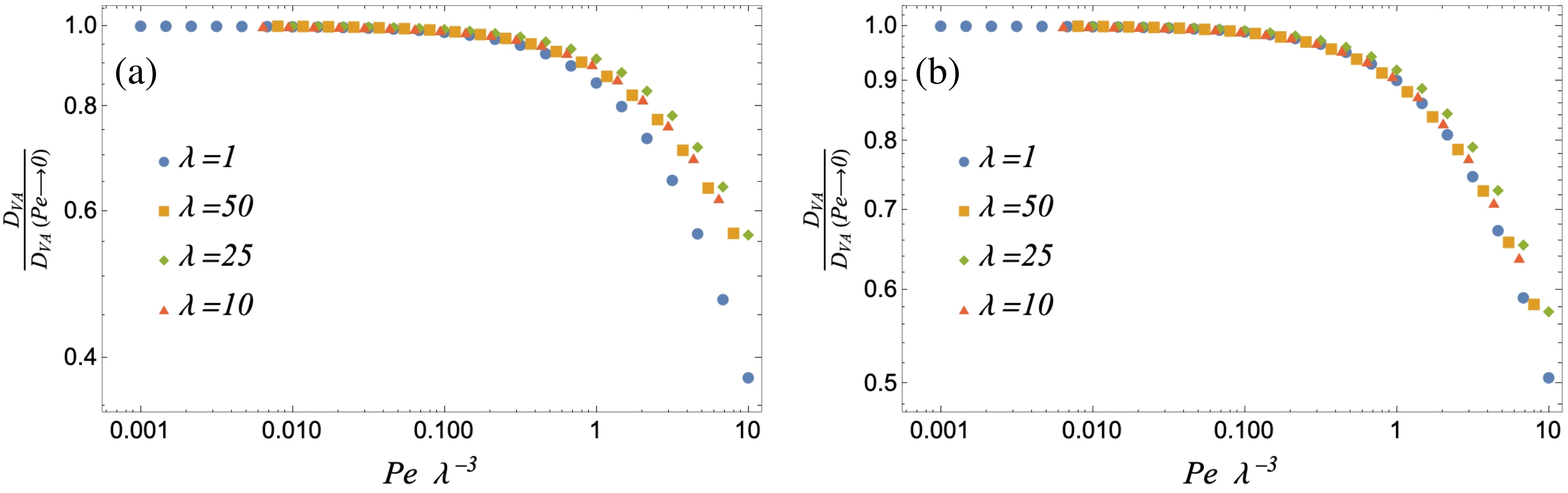}
\caption{Onsager cross coupling coefficient $D_{VA}$, computed using numerical simulations of the governing equations linearized around the equilibrium.  In panel (a) we show the case of $Da=1$ and in panel (b) the case of $Da=2$. The dimensionless numbers are $\beta=1$ and  $\epsilon=1$. The data computed at different interaction potential range, $\lambda$, collapse on a mastercurve up to  $Pe \, \lambda^{-3} \approx 1$.}
\label{fig5}
\end{figure}

In Figure \ref{fig4}, we show the Onsager cross-coupling coefficients for values of $\epsilon$ and $Da$ that are beyond the range of applicability of the perturbation expansion. The numerical results show that $D_{VA}=D_{WF}$ for all the parameters investigated, thus confirming that the Onsager's reciprocal relations are fulfilled by the governing equations even beyond the range of applicability of the perturbation expansion. Interestingly, in the limit $Pe\rightarrow 0$, the cross-coupling coefficients attain a constant value that is independent of $Pe$ and depends only on $\epsilon$ and $Da$. The range of $Pe$ for which the $D_{VA}$ and $D_{WF}$ are constant depends on the range of the interaction potential $\lambda^{-1}$. For short-ranged potentials, $\lambda^{-1} \gg 1$, the coupling coefficients are constant up to very large values of $Pe$. To investigate the effect of the range of the interaction potential, $\lambda^{-1}$, in Figure \ref{fig5} we plot the cross-coupling coefficients, normalized by their value at $Pe \rightarrow 0$, as a function of $Pe \, \lambda^{-3}$. The results show that $D_{VA}$ and $D_{WF}$ calculated for different interaction ranges, $\lambda^{-1}$, collapse onto a mastercurve that depends only on $\epsilon$ and $Da$. For $Pe \, \lambda^{-3} \ll 1$ the cross-coupling coefficients are constant and they start to decay to zero when $Pe \, \lambda^{-3} \approx 1$. This scaling is in agreement with the findings of Michelin and Lauga \cite{michelin2014phoretic} who found that in the limit $\lambda^{-1} \gg 1$, the advection of species becomes important within the thin boundary layer only if $Pe \approx \lambda^{3}$. Our numerical simulations suggest that,  for $Pe \, \lambda^{-3} \ll 1$,  advection can be safely neglected if one is interested in the propulsion of chemically-active colloids. However, one should retain advection in cases where external forces are present, since neglecting it leads to $D_{WF}=0$ thus breaking the Onsager reciprocal relations. 

Our results suggest that the momentum balance and the transport of solute are coupled even in the limit of $Pe\rightarrow 0$.  Such coupling is necessary for an external force to drive a chemical reaction and preserve the symmetry of the Onsager relations. Indeed, the Force balance, given by Eq. \eqref{force_bal_dim}, reveals that the velocity field must scale as $v \propto F^*/Pe$ in the limit of $Pe\rightarrow 0$. By substituting this scaling into the transport equation of the species B, the $Pe$ number that multiplies in the advective term of the equation cancels out with the scaling $v \propto F^*/Pe$.
From a physical standpoint, in the limit $Pe\rightarrow 0$, the phoretic velocity scale used in the definition of the P\'eclet number becomes irrelevant and the only relevant velocity scale can be constructed using the external force $F$. One can redefine a new P\'eclet number using this velocity scale, which would contain the mechanical affinity in its definition. The immediate consequence of this is that one cannot simultaneously consider a finite mechanical affinity and vanishing advective effects.


Our results are in agreement with the recent work by Gaspard and Kapral  \cite{gaspard2018nonequilibrium}, who propose that in the limit of short-range potentials, there is a coupling between the tangential component of the traction exerted by the fluid and the tangential transport of species.  Such coupling is independent of the P\'eclet number and couples the transport of solute and the transport of momentum even if the advective transport outside the boundary layer is negligible.  

\section{Onsager reciprocal relations around a nonequilibrium steady state}\label{sec7}
In this section we investigate the validity of the Onsager reciprocal relations around a nonequilibrium steady state. We use the finite element method to solve the base state, given by Eqs.\eqref{dimensionlessmombal}-\eqref{force_bal_dim}, and compute the steady-state quantities. We assume that the base state is given by an active particle driven by an external force $F^*_0$ or by a chemical affinity $A_{\text{rxn},0}$. As we did in the previous section, we fix the chemical activity as $g(\theta)= 1+\cos{(\theta)}$ and the interaction potential as $f(r,\theta)= \epsilon \exp{(\lambda (r-1))}\left[\cos{(\theta})-1\right]$.  The nonlinear system of equations is solved using the Newton-Raphson method starting from an initial guess given by the equilibrium distribution of species. Once the base state is computed, we solve the linearized equations, Eqs.\eqref{lindimensionlessmombal}-\eqref{lindimensionlesscbboundcond}, using the same mesh used to solve for the base state. 

\begin{figure}[h!]
\centering
\includegraphics[width=1.0\textwidth]{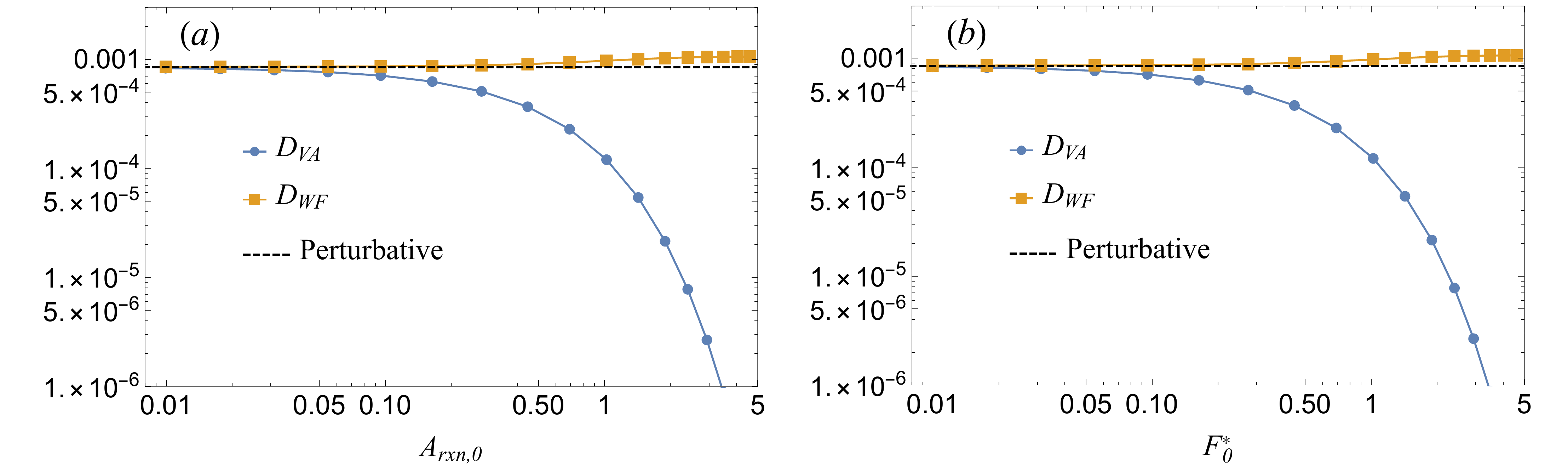}
\caption{Onsager cross-coupling coefficients, $D_{WF}$ and $D_{VA}$, computed using numerical simulations of the governing equations linearized around a nonequilibrium steady state. In the panel (a) the base state is driven out of equilibrium by a nonzero chemical affinity $A_{\text{rxn},0}$, while in panel (b) the base state is driven out of equilibrium by an external force $F_{0}^*$. The remaining dimensionless numbers are $\beta=Pe=1$, $\epsilon = 0.1$ and $Da=0.1$.}
\label{fig3}
\end{figure}

In Figure \ref{fig3}, we report $D_{WF}$ and $D_{VA}$ for a base state driven out of equilibrium by an external force or by the chemical affinity for the case $Pe=\beta=1$, $\epsilon = 0.1$ and $Da=0.1$. In panel (a) and in panel (b) of Figure \ref{fig3} it is apparent that for small thermodynamic forces the system is near equilibrium and $D_{WF}=D_{VA}$ with their value agreeing with the asymptotic approximation given by Eq. \eqref{xi11} and Eq. \eqref{finaleqreac1}. However, Figure \ref{fig3} shows that far from equilibrium the two coefficients are different meaning that the Onsager reciprocal relations break down. Here, we also find that considering a generalized chemical affinity as proposed in \cite{pagonabarraga1997fluctuating,bedeaux2011concentration} does not restore the symmetry of the Onsager relations. Since in experimental conditions the active particles are usually driven by a chemical reaction that is far from equilibrium, we expect the Onsager reciprocal relations to be broken. 

\section{Conclusions}\label{sec8}
We investigate the Onsager reciprocal relations for a chemically-active colloid. We assume that the active colloid is suspended in an incompressible solution of two species, A and B, with the species B interacting through a potential with the surface of a spherical particle. The two species undergo a reversible reaction at the surface of the colloid. In the case of the thermodynamic system investigated here, the Onsager reciprocal relations link the total surface reaction rate and the velocity of the active colloid to the chemical and the mechanical affinity. Such chemo-mechanical coupling can be formalized using the Onsager matrix, which must be symmetric positive definite around the equilibrium.

Here we derive the Onsager reciprocal relations, starting from the local transport equations of the number density of species, the balance of momentum, and the continuity equation. These equations are defined in the volume outside the active colloid and are derived using the framework of nonequilibrium thermodynamics and the assumption of local equilibrium. Since the resulting governing equations are nonlinear, we linearize them around a generic steady state. Using a perturbation expansion and numerical simulations we compute the Onsager matrix. We show that the Onsager reciprocal relations are recovered when the equations are linearized around the thermodynamic equilibrium. This is expected since at equilibrium the microscopic equations of motion obey the detailed balance. In addition, our results agree with the self-phoretic velocity calculated in previous works using matched asymptotic expansions  \cite{sabass2012dynamics}. We find that accounting for the advection of the reacting species is crucial to preserve the symmetry of the Onsager matrix even in the case of short-ranged interaction potentials or vanishing P\'eclet numbers. Neglecting the advective transport of the solute breaks the symmetry of the Onsager relations. In the limit of vanishing P\'eclet numbers, the only relevant velocity scale can be defined using the mechanical affinity. As a consequence, the mechanical affinity enters the definition of the P\'eclet number and one cannot simultaneously neglect the advective transport of the solutes and consider a finite mechanical affinity: A nonzero mechanical affinity implies nonzero advective effects.


Finally, we investigated the validity of the Onsager reciprocal relations around a nonequilibrium steady state (NESS). The active particle is driven by an external force or by a nonzero chemical affinity and we consider small perturbations around this nonequilibrium steady state. Previous works have shown that the reciprocal relations might hold around NESS even if the detailed balance of the underlying dynamics is broken \cite{gabrielli1996onsager,gabrielli1999onsager,dal2019linear}. Here, we found that the symmetry of the Onsager reciprocal relations breaks down and one cannot define an effective temperature that preserves the symmetry of the Onsager matrix \cite{Hargus_2021,JClub_Grosberg}. 
Indeed, most of the active particles used in experiments are driven far from equilibrium and we should expect their Brownian motion to be qualitatively different from that experienced at equilibrium \cite{Golestanian_2009,Gomez_Solano_2016}.

\begin{acknowledgments}
M.D.C. acknowledges funding from the European Union’s Horizon 2020 research and innovation program under the Marie Skłodowska-Curie action (GA 712754), the Severo Ochoa programme (SEV-2014-0425), the CERCA Programme/Generalitat de Catalunya, and the Spanish Ministry of Science and Innovation (MCIN) under the Juan de la Cierva (IJC2018-035270-I) postdoctoral grant and the retos de investigacion project PID2020-113033GB-I00. I.P. acknowledges support from MINECO/FEDER Project No. PGC2018-098373-B-I00, DURSI Project No. 2017SGR-884, SNF Project No. 200021-175719. M.D.C and I.P ackowledge funding from the H2020 research and innovation program under the FET open project NanoPhlow (GA 766972).
\end{acknowledgments}

\bibliography{Onsager_matrix_bib.bib}

\end{document}